\documentclass[prd,twocolumn]{revtex4}
\usepackage{graphicx, epsfig}
\usepackage{color}
\usepackage{mathrsfs}
\usepackage{amsmath}

\newcommand{\be}{\begin{equation}}
\newcommand{\ee}{\end{equation}}
\newcommand{\bea}{\begin{eqnarray}}
\newcommand{\eea}{\end{eqnarray}}

\newcommand{\gapp}{\mathrel{\raise.3ex\hbox{$>$}\mkern-14mu
              \lower0.6ex\hbox{$\sim$}}}
\newcommand{\lapp}{\mathrel{\raise.3ex\hbox{$<$}\mkern-14mu
              \lower0.6ex\hbox{$\sim$}}}

\begin{document}

\title{Quantum gravitational collapse: non-singularity and non-locality}
\author{Eric Greenwood}
\author{Dejan Stojkovic}
\affiliation{HEPCOS, Department of Physics,
SUNY at Buffalo, Buffalo, NY 14260-1500}
 %%%%%%%%%%%%%%%%%%%%%%%%%%%%%%%%%%%%%%%%%%%%%%%%%%%%%%%
\begin{abstract}
We investigate gravitational collapse in the context of quantum mechanics. We take primary interest in the behavior of the collapse near the horizon and near the origin (classical singularity) from the point of view of an infalling observer. In the absence of radiation, quantum effects near the horizon do not change the classical conclusions for an infalling observer, meaning the horizon is not an obstacle for him. However, quantum effects are able to remove the classical singularity at the origin, since the wave function is non-singular at the origin. Also, near the classical singularity, some non-local effects become important. In the Schrodinger equation describing behavior near the origin, derivatives of the wave function at one point are related to the value of the wave function at some other distant point.
\end{abstract}

%\pacs{04.50.Gh, 04.70.Dy}

\maketitle

\section{Introduction}
Of the four fundamental interactions in nature, gravity is by far
the weakest.  For this reason, we can hope to see quantum effects
only in the vicinity of classical singularities.  Penrose and
Hawking \cite{PH} have shown that these singularities are endemic in general
relativity. However, the question arises whether they are an
intrinsic property of space-time or simply reflect our lack of the
ultimate non-singular theory.  A singularity often represents a
signal that  the theory has been extrapolated outside its domain
of validity. The general belief is that quantization will rid
gravitation of singularities, just as in atomic physics it got rid
of the singularity of the Coulomb potential which has an identical $1/r$ behavior \cite{Bogojevic:1998ma,Trodden:1993dm,MankocBorstnik:2005ib,Shankaranarayanan:2003qm}.  If this is indeed the case, this information must be encoded in the wave function describing the collapsing object.
There are also some reasons to believe that inherently quantum  non-local effects should become important in the strong gravity regime, i.e. near the classical singularity. These effects might be associated with the possible resolution of the information loss paradox \cite{Lowe:1999pk,Horowitz:2003he,Giddings:2006sj,Giddings:2006be}.

There are additional reasons to study quantum effects in gravitational collapse even in the regions far from the origin.
Properties of the classical Schwarzschild solution are very well understood. If an asymptotic observer throws something down a pre-existing black hole, he will never see it crossing the horizon since it takes infinite time $\Delta t$ according to his clock. If $R_0$ is the position of a fixed outside observer then
\be
\Delta t = \int_{R_S}^{R_0} \frac{dr}{1-\frac{R_S}{r}} \rightarrow \infty \, .
\ee
The integral is obviously divergent at the lower limit.
For the same reason,  an asymptotic observer will never see the formation of a black hole.  Thus, an asymptotic observer will never see any effects strictly associated
with an event horizon: all of such signals (light rays or gravitational waves) will be infinitely redshifted. Thus, one can study the effects arising in near horizon region but could never probe the horizon itself. This includes black holes in the centers of galaxies, black hole mergers, accretion of material onto black holes, x-ray bursts etc. This fact is often ignored. The hope is that quantum mechanics will make the time finite even for an asymptotic observer.

How can quantum mechanics be important for large macroscopic black holes? If one includes quantum fluctuations, the position of a horizon is not fixed. Then, instead of $R_S$ one can write  $R_S + \delta R_S$, where $\delta R_S$ represent small fluctuations in the position of a horizon. This can make the time as measured by an asymptotic observer finite (see e.g. Sec.~10.1.5 of \cite{frolov})):
\be
\Delta t = \int_{R_S + \delta R_S}^{R_0} \frac{dr}{1-\frac{R_S}{r}} \sim R_S \ln \left( \frac{R_0 - R_S}{\delta R_S} \right)\, .
\ee
 If this were true, it would have a profound importance for astrophysical observations. We would be able to observe black hole formation
and other effects in finite time.  Note however that fluctuations can go either way, i.e.
$R_S$ should get replaced by $R_S \pm \delta R_S$. In the case of $R_S - \delta R_S$, the result becomes infinite again. This points out that a careful analysis of gravitational collapse in the context of quantum mechanics is needed. This question was investigated in \cite{VSK}, from the point of view of an asymptotic observer. The conclusion was (with some caveats) that an asymptotic observer does
not see a collapsing object crossing its own Schwarzschild radius even when quantum effects are taken into account.

In this paper we address the question of gravitational collapse in the context of quantum mechanics as seen by an infalling observer. We model the collapsing body with a shell which is effectively described by a spherically symmetric domain wall. The reason for that is the existence of a well defined relativistic Lagrangian and Hamiltonian
for such a system which can be quantized straightforwardly.

In Sec.~\ref{setup} we adopt the Wheeler-de Witt approach to quantum gravity and define the framework. For all  practical purposes in our case of interest, this formalism is equivalent to the
functional Schrodinger formalism.
In Sec.~\ref{asymptotic} we review the recent results obtained in \cite{VSK} for an asymptotic observer. In Sec.~\ref{classical_in} we study classical collapse from the point of view of an infalling observer (an observer who is falling together with the collapsing shell). The most important results are presented in Sec.~\ref{qcollapse_in} where we study collapse from the point of view of an infalling observer in the context of quantum mechanics. In Sec.~\ref{near_horizon} we explore quantum effects in near horizon limit for an infalling observer and show that, in the absence of quantum radiation, classical conclusions remain true, i.e. horizon is no obstacle for an infalling observer. In Sec.~\ref{near_singularity} we explore quantum effects near the origin (i.e. classical singularity) and demonstrate two results: a) the wave function describing the collapsing shell is non-singular at the origin, and b) the non-local effects, which were absent at large distances, become unsuppressed in this near the origin regime.

We emphasis that we do not study quantum radiation of the fields propagating in the background of a collapsing object which may introduce some new elements. [This question will be studied in \cite{Greenwood:2008zg}.] Our conclusions are summarized in Sec.~\ref{conclusions}.

\section{Setup and formalism}
\label{setup}

The Wheeler-de Witt equation \cite{DeWitt} in its general form is
\begin{equation}
H \Psi = 0
\end{equation}
where $H$ is the Hamiltonian and $\Psi [X^\alpha,g_{\mu\nu},{\cal O}]$ is the wave-functional for all the ingredients of the system - collapsing object, space-time metric, even the observer's degrees of freedom (denoted by ${\cal O}$). We will separate the Hamiltonian into two parts, one for the system and the other for the observer
\begin{equation}
H = H_{\rm sys} + H_{\rm obs}
\end{equation}
Any weak interaction terms between the observer and the system are included in $H_{\rm sys}$.
The observer is assumed not to significantly affect the evolution of the system. The total wave-functional
can be written as a sum over eigenstates
\begin{equation}
\Psi = \sum_k c _k \Psi^k_{\rm sys} ({\rm sys},t) \Psi^k_{\rm obs} ({\cal O},t)
\end{equation}
where $k$ labels the eigenstates, $c_k$ are complex coefficients, and we have introduced the observer time, $t$, via
\begin{equation}
i \frac{\partial \Psi^k_{\rm obs}}{\partial t}
           \equiv H_{\rm obs} \Psi^k_{\rm obs}
\end{equation}
With the help of an integration by parts, and the fact that the total wave-functional is independent of $t$, the Wheeler-de Witt equation implies the Schrodinger equation
\begin{equation}
H_{\rm sys} \Psi^k_{\rm sys} =
i \frac{\partial \Psi^k_{\rm sys}}{\partial t}
\label{schrodingerpre}
\end{equation}
For convenience, from now on we will denote the system wave-function
simply by $\Psi$ and drop the superscript $k$ and the subscript
``${\rm sys}$''. Similarly $H$ will now denote $H_{\rm sys}$, and the Schrodinger equation reads
\begin{equation}
H \Psi = i \frac{\partial \Psi}{\partial t}
\label{schrodinger}
\end{equation}

\section{Asymptotic observer}
\label{asymptotic}

A general treatment of the full Wheeler-de Witt equation is very
difficult and we truncate the field degrees of freedom to a finite subset. In other words, we will consider a minisuperspace version of the Wheeler-de Witt equation. In particular,
we only consider spherical domain wall representing spherical shell of collapsing matter. This implies that the wall is described by only the radial degree of freedom, $R(t)$. The metric is taken to be the solution of Einstein equations for a spherical domain wall.
The metric is Schwarzschild outside the wall, as follows
from spherical symmetry \cite{Ipser:1983db}
\begin{equation}
ds^2= -(1-\frac{R_S}{r}) dt^2 + (1-\frac{R_S}{r})^{-1} dr^2 +
      r^2 d\Omega^2 \ , \ \ r > R(t)
\label{metricexterior}
\end{equation}
where, $R_S = 2GM$ is the Schwarzschild radius in terms of the mass,
$M$, of the wall, and
\begin{equation}
d\Omega^2  = d\theta^2  + \sin^2\theta d\phi^2
\end{equation}
In the interior of the spherical domain wall, the line element
is flat, as expected by Birkhoff's theorem,
\begin{equation}
ds^2= -dT^2 +  dr^2 + r^2 d\theta^2  + r^2 \sin^2\theta d\phi^2  \ ,
\ \ r < R(t)
\label{metricinterior}
\end{equation}
The equation of the wall is $r=R(t)$.
The interior time coordinate, $T$, is related to the asymptotic observer time coordinate, $t$, via the proper time of an observer moving with the shell, $\tau$. The relations are
\begin{equation}
\frac{dT}{d\tau} =
      \left [ 1 + \left (\frac{dR}{d\tau} \right )^2 \right ]^{1/2}
\label{bigTandtau}
\end{equation}
and
\begin{equation}
\frac{dt}{d\tau} = \frac{1}{B} \left [ B +
         \left ( \frac{dR}{d\tau} \right )^2 \right ]^{1/2}
\label{littletandtau}
\end{equation}
where
\begin{equation}
B \equiv 1 - \frac{R_S}{R}
\label{BofR}
\end{equation}

By integrating the equations of motion for the spherical domain wall, Ipser and Sikivie \cite{Ipser:1983db} found that the mass is a constant of motion and is given by
\begin{equation}
M = \frac{1}{2} [ \sqrt{1+R_\tau^2} + \sqrt{B+ R_\tau^2} ] 4\pi \sigma R^2
\label{ISmass}
\end{equation}
where $R_{\tau} = dR/d\tau$, while $\sigma$ is the surface tension (energy density per unit area) of the wall. It is assumed that ${\rm max} (R) < (4\pi G\sigma )^{-1} $ to avoid the case in which the domain wall is already within its own Schwarzschild radius to begin with. This expression for $M$ is implicit since $R_S =2GM$ occurs in $B$.
Solving for $M$ explicitly in terms of $R_\tau$ gives
\begin{equation}
M = 4\pi \sigma R^2 [ \sqrt{1+R_\tau^2} - 2\pi G\sigma R] .
\label{MRtau}
\end{equation}

\subsection{Classical collapse as viewed by an asymptotic observer}
\label{classical_as}

The effective Lagrangian describing gravitational collapse of the spherical shell consistent with (\ref{MRtau}) was derived in \cite{VSK}

\begin{equation}
L_{eff} = - 4\pi \sigma R^2
        \left [ \sqrt{B-\frac{{\dot R}^2}{B}} - 2\pi G \sigma R
                \sqrt{B - \frac{1-B}{B} {\dot R}^2} \right ]
\label{efflagrangian}
\end{equation}
where ${\dot R} = dR/dt$.
The generalized momentum, $\Pi$, can be derived from
Eq.~(\ref{efflagrangian})
\begin{equation}
\Pi = \frac{4\pi \sigma R^2 {\dot R}}{\sqrt{B}} \left [
      \frac{1}{\sqrt{B^2-{\dot R}^2}} -
       \frac{2\pi G\sigma R (1-B)}{\sqrt{B^2 - (1-B) {\dot R}^2}}
                        \right ]
\label{momentumA}
\end{equation}
The Hamiltonian (in terms of ${\dot R}$) is
\begin{equation}
H = 4\pi \sigma B^{3/2}R^2 \left [
         \frac{1}{\sqrt{B^2-{\dot R}^2}} -
          \frac{2\pi G\sigma R}{\sqrt{B^2- (1-B){\dot R}^2}}
                                  \right ]
\label{Ham}
\end{equation}

To obtain $H$ as a function of $(R, \Pi )$, we need to eliminate
${\dot R}$ in favor of $\Pi$ using Eq.~(\ref{momentumA}). This can
be done but is messy, requiring solutions of a quartic polynomial.
Instead we consider the limit when $R$ is
close to $R_S$ and hence $B \rightarrow 0$. In this limit the
denominators of the two terms in Eq. (\ref{Ham}) are equal and
\begin{equation}
\Pi \approx \frac{4\pi \mu R^2  {\dot R}}
              {\sqrt{B} \sqrt{B^2-{\dot R}^2}}
\end{equation}
where
\begin{equation}
\mu \equiv \sigma (1-2\pi G\sigma R_S)
\end{equation}
and
\begin{eqnarray}
H &\approx& \frac{4\pi \mu B^{3/2}R^2}{\sqrt{B^2-{\dot R}^2}} \label{HRdot}\\
  &=& \left [  (B\Pi)^2 + B (4\pi \mu R^2)^2 \right ] ^{1/2} \label{HPi}
\end{eqnarray}
The Hamiltonian has the form of the energy of a relativistic
particle, $\sqrt{p^2 + m^2}$, with a position dependent mass.

The Hamiltonian is a conserved quantity and so, from Eq.~(\ref{HRdot}),
\begin{equation}\label{defh}
\frac{ B^{3/2}R^2}{\sqrt{B^2-{\dot R}^2}} =h
\end{equation}
where $h = H/4\pi \mu $ is a constant. (Up to the approximation
used to obtain the simpler form of the Hamiltonian in Eq.~(\ref{HRdot}),
the Hamiltonian is the conserved mass.)

Solving Eq. (\ref{defh}) for ${\dot R}$ we get
\begin{equation}
{\dot R} = \pm B \left(1- \frac{BR^4}{h^2} \right)^{1/2}\, ,
\label{Rdotsolution}
\end{equation}
which, near the horizon, takes the form
\begin{equation}
{\dot R} \approx \pm B\left(1- {1 \over 2} \frac{BR^4}{h^2} \right)
\label{rdotnh}
\end{equation}
since $B \rightarrow 0$ as $R \rightarrow R_S$.

The dynamics for $R \sim R_S$ can be obtained by solving the equation
${\dot R} =  \pm B$. To leading order in $R-R_S$, the solution is
\begin{equation}
R(t) \approx R_S + (R_0-R_S) e^{\pm t/R_S} \, .
\label{solution}
\end{equation}
where $R_0$ is the radius of the shell at $t=0$. As we are interested
in the collapsing solution, we choose the negative sign in the exponent.
This solution implies that, from the classical point
of view, the asymptotic observer never sees the formation of the horizon
of the black hole, since $R(t) = R_S$ only as $t \rightarrow \infty$.
Thus, the time needed for a collapsing object to cross its own Schwarzschild radius is infinite from a point of view of a static outside observer.

\subsection{Quantum collapse as viewed by an asymptotic observer}
\label{qcollapse_as}

The classical Hamiltonian in Eq.~(\ref{HPi}) has a square root and
so we first consider the squared Hamiltonian
\begin{equation}\label{Hsq}
H^2 = B\Pi ~B \Pi + B (4\pi \mu R^2)^2
\end{equation}
where we have made a choice for ordering $B$ and $\Pi$ in
the first term. Now we apply the standard quantization procedure. We substitute
\begin{equation}
\Pi = -i \frac{\partial}{\partial R}
\end{equation}
in the squared Schrodinger equation,
\begin{equation}
H^2 \Psi = - \frac{\partial^2 \Psi}{\partial t^2}
\label{Hsquared}
\end{equation}
Then
\begin{equation}\label{Hsqexplicit}
- B\frac{\partial}{\partial R}
 \left ( B \frac{\partial \Psi}{\partial R} \right ) +
    B (4\pi \mu R^2)^2 \Psi = - \frac{\partial^2 \Psi}{\partial t^2}
\end{equation}
To solve this equation, define
\begin{equation}
u = R + R_S \ln \left | \frac{R}{R_S} - 1 \right |
\label{uandR}
\end{equation}
which gives
\begin{equation}
B\Pi = -i \frac{\partial}{\partial u}
\end{equation}
Eq.~(\ref{Hsquared}) then gives
\begin{equation}
  \frac{\partial^2 \Psi}{\partial t^2}
 - \frac{\partial^2 \Psi}{\partial u^2}
 + B (4\pi \mu R^2)^2 \Psi = 0
\label{waveeq}
\end{equation}
This is just the massive wave equation in a Minkowski background
with a mass that depends on the position. Note that $R$ needs to
be written in terms of the coordinate $u$ and this can be done
(in principle)
by inverting Eq.~(\ref{uandR}). However, care needs to be taken
to choose the correct branch since the region
$R \in (R_S,\infty)$ maps onto $u\in (-\infty ,+\infty)$
and $R \in (0,R_S)$ onto $u \in (0,-\infty)$.

We are interested in the situation of a collapsing shell.
In the region $R \sim R_S$, the logarithm in Eq.~(\ref{uandR})
dominates and
$$
R \sim R_S + R_S e^{u/R_S}
$$
We look for wave-packet solutions propagating toward
$R_S$, that is, toward $u \rightarrow - \infty$. In this limit
$$
B \sim e^{u/R_S} \rightarrow 0
$$
and the last term in Eq.~(\ref{waveeq}) can be ignored.
Wave packet dynamics in this region is simply given by the
free wave equation and any function of light-cone coordinates
($u\pm t$) is a solution. In particular, we can write a Gaussian
wave packet solution that is propagating toward the Schwarzschild
radius
\begin{equation}
\Psi = \frac{1}{\sqrt{2\pi} s} e^{- (u+t)^2/2s^2 }
\label{packetsolution}
\end{equation}
where $s$ is some chosen width of the wave packet in the $u$
coordinate. The width of the Gaussian wave packet remains fixed in
the $u$ coordinate while it shrinks in the $R$ coordinate via the
relation $dR = B du$ which follows from Eq.~(\ref{uandR}).   This fact
is of great importance, since if the wave packet remained of constant size
in $R$ coordinates, it might cross the event horizon in finite time.

The wave packet travels at the speed of light in the $u$ coordinate
-- as can be seen directly from the wave equation Eq.~(\ref{waveeq})
or from the solution, Eq.~(\ref{packetsolution}). However, to get to the
horizon, it must travel out to $u=-\infty$, and this takes an infinite
time. So we conclude that the quantum domain wall does not collapse
to $R_S$ in a finite time, as far as the asymptotic observer is concerned,
so that quantum effects considered here do not alter the classical result that
an asymptotic observer does not observe the formation of an event
horizon.

This analysis leaves room for a non-local process by which
the wave packet can jump from the $(R_S,\infty)$ region to the $(0,R_S)$ region, without ever going through $R_S$, since the region
$R \in (R_S,\infty)$ maps onto $u\in (-\infty ,+\infty)$
and $R \in (0,R_S)$ onto $u \in (0,-\infty)$. Note that this process would be quite different from tunneling through a barrier.

We end this section with the remark that quantum Hawking
emission makes the lifetime of the black hole finite.  Therefore,  asymptotic observers  will not see infalling matter hovering over the horizon for an infinite time. From the point of view of outside
observers, the evaporation process and the resulting semi-classical geometry can be described in terms of a stretched horizon \cite{Price:1986yy,Susskind:1993if} rather than focusing on the event horizon. In this simple physical picture  infalling matter never passes through the event horizon but is instead absorbed into the stretched horizon, thermalized, and eventually re-emitted as part of the Hawking radiation.

\section{Classical collapse as viewed by an infalling observer}
\label{classical_in}

We now change the observer and study what an observer who is falling together with the shell (i.e. an infalling observer) would see.
The effective action consistent with the conserved quantity (\ref{MRtau}) is
\begin{equation}
  S_{eff}=-4\pi\sigma\int d\tau R^2\left[\sqrt{1+R_{\tau}^2}-R_{\tau}\sinh^{-1}(R_{\tau})-2\pi\sigma GR\right].
\end{equation}
Therefore the effective Lagrangian expressed in terms of the infalling observer's time $\tau$ is given by
\begin{equation}
  L_{eff}=-4\pi\sigma R^2\left[\sqrt{1+R_{\tau}^2}-R_{\tau}\sinh^{-1}(R_{\tau})-2\pi\sigma GR\right].
  \label{Lagrangian}
\end{equation}

The generalized momentum, $\Pi$, can be derived from Eq. (\ref{Lagrangian})
\begin{equation}
  \Pi=4\pi\sigma R^2\sinh^{-1}(R_{\tau}).
  \label{momentum}
\end{equation}
The Hamiltonian (in terms of $R_{\tau}$) is
\begin{equation}
  H=4\pi\sigma R^2\left[\sqrt{1+R_{\tau}^2}-2\pi\sigma GR\right]
  \label{full Ham}
\end{equation}
which is just  Eq. (\ref{MRtau}). Let us briefly comment on the physical meaning of the Hamiltonian (\ref{full Ham}). For a static shell, i.e. $R_{\tau} =0$, the first term in square brackets is just the total rest mass of the shell. For a moving shell, $R_{\tau} \neq 0$ takes kinetic energy into account. The last term in square brackets is the self-gravity, since in this formalism the collapsing shell is both the source of the gravitational field and the matter that is collapsing.

 From (\ref{full Ham}) we can calculate $R_\tau$
\be \label{rtau}
R_\tau = \pm \sqrt{\left( \frac{h}{R^2} + 2\pi \sigma GR \right)^2-1}
\ee
where $h=H/(4 \pi \sigma )$. In our formalism, the Hamiltonian is just the conserved mass, $H =M$ (the value of which can be viewed as an initial condition).
When the shell comes close to its own Schwarzschild radius, i.e. $R \approx R_S$, this equation can be further simplified. If we are interested only in the zeroth order behavior near the Schwarzschild radius we can set $R = R_S$ in (\ref{rtau}). This gives us
\be
R_\tau = \pm \sqrt{\left( \frac{h}{R_S^2} + 2\pi \sigma GR_S \right)^2-1}
\ee
which can be integrated to give
\be
R  = R_0 - \tau \sqrt{\left( \frac{h}{R_S^2} + 2\pi \sigma GR_S \right)^2-1}
\ee
where $R_0$ is the radius of the shell at $\tau=0$. The solution implies that the infalling observer will reach $R_S$ in a finite amount of his proper time, which is expected from classical general relativity.

\section{Quantum collapse as viewed by an infalling observer}
\label{qcollapse_in}

\subsection{Quantum treatment of gravitational collapse near the horizon}
\label{near_horizon}

Now we turn to quantum treatment of gravitational collapse as viewed by an infalling observer. The exact Hamiltonian (in terms of $R_{\tau}$) is again (\ref{full Ham})
\begin{equation}
  H=4\pi\sigma R^2\left[\sqrt{1+R_{\tau}^2}-2\pi\sigma GR\right]
  \end{equation}
In this section, we will quantize this Hamiltonian near the Schwarzschild radius and in the limit where $R_\tau$ is small.
This is indeed a restriction to the special motion of the wall, since in general $R_\tau$ can be large near $R_S$ if the shell is falling from a very large distance. However, one may always choose initial conditions in such a way that the initial position of the shell $R(\tau =0)$ is very close to $R_S$.

In the limit of small $R_\tau$ and $R \rightarrow R_S$  the Hamiltonian simplifies to
\be
H=4\pi\sigma R_S^2\left[1+\frac{1}{2} R_{\tau}^2-2\pi\sigma GR_S\right]
\ee
In the same limit, the momentum (\ref{momentum}) simplifies to
\be \label{appmom}
\Pi =4\pi \sigma R_S^2 R_\tau
\ee
Dropping the constant terms form the Hamiltonian we get
\be
H=\frac{\Pi^2}{8\pi\sigma R_S^2} .
\ee
Using the standard quantization procedure, we substitute
\begin{equation}
  \Pi=-i\frac{\partial}{\partial R}
  \label{pi}
\end{equation}
into the Schr\"{o}dinger equation,
\begin{equation}
  H\Psi=i\frac{\partial\Psi}{\partial \tau}.
  \label{Quant Ham}
\end{equation}
This gives us
\begin{equation}
  -\frac{1}{8\pi\sigma R_S^2} \frac{\partial^2 \Psi}{\partial R^2} = i\frac{\partial\Psi}{\partial\tau} .
\end{equation}
This is just the Schrodinger equation for a freely propagating "particle" of mass $4 \pi\sigma R_S^2$, as we expected in this approximation. Since $R_S$ is only a finite distance away for an
infalling observer we conclude that quantum effects do not alter the classical result - a collapsing shell crosses its own Schwarzschild radius in a finite proper time.

\subsection{Quantum treatment of gravitational collapse near the origin}
\label{near_singularity}

In this section we investigate the most important question of quantum effects when the collapsing shell approaches the origin (i.e. classical singularity at $R\rightarrow0$). The exact Hamiltonian (in terms of $R_{\tau}$) is again
\begin{equation}
  H=4\pi\sigma R^2\left[\sqrt{1+R_{\tau}^2}-2\pi\sigma GR\right]
\end{equation}
while exact value of $R_\tau$ is
\be
R_\tau = \pm \sqrt{\left( \frac{h}{R^2} + 2\pi \sigma GR \right)^2-1}
\ee	
Near the origin, i.e. in the limit of $R\rightarrow 0$ the classical solution for $R_\tau $  (keeping only the leading order term) becomes
\be \label{ras}
  R_{\tau} \approx-\frac{h}{R^2} .
\ee
where $h=H/(4 \pi \sigma )$, while the Hamiltonian is just the conserved mass $H =M$.
This implies that, up to the leading term  near the origin, the Hamiltonian is
\begin{equation}
  H=4\pi\sigma R^2 R_{\tau} .
\end{equation}
Substituting the asymptotic behavior (\ref{ras}) in the expression for the generalized momentum
\be
  \Pi=4\pi\sigma R^2\sinh^{-1}(R_{\tau}).
\ee
we learn that
\be
 \lim_{R\rightarrow 0} \, \Pi=0
\ee
and
\be
 \lim_{R\rightarrow 0} \, \frac{\Pi}{4\pi\sigma R^2}=- \infty .
\ee
This implies that $R_\tau $ defined as
\be
R_\tau = \sinh \left(\frac{\Pi}{4\pi\sigma R^2} \right),
\ee
near the origin becomes
\be \label{no}
R_\tau = \frac{1}{2} \exp \left(-\frac{\Pi}{4\pi\sigma R^2}\right) .
\ee
Therefore we can write the Schr\"{o}dinger equation as
\begin{equation}
  2\pi\sigma R^2\exp\left(\frac{i}{4\pi\sigma R^2}\frac{\partial}{\partial R}\right)\Psi (R, \tau)=i\frac{\partial\Psi (R, \tau)}{\partial\tau}.
  \label{near origin}
\end{equation}

In Eq. (\ref{near origin}), the differential operator in the exponent gives some unusual properties to the equation. Note that if we expand the exponent we can not stop the series after the finite number of terms but need to include all of the terms. This means that we need to include an infinite number of derivatives of the wave function $\Psi$ into the differential equation. An infinite number of derivatives of a certain function uniquely specifies the whole function. Thus, the value of (the derivative of) the function on the right hand side of
Eq. (\ref{near origin}) at one point depends on the values of the function at different points on the left hand side of the same equation. This is in strong contrast with ordinary local differential equations where the value of the function and certain finite number of its derivatives are related at the same point of space.
This indicates that Eq. (\ref{non-local}) describes physics which is not strictly local.

With the simple change of variables we can make this argument more transparent. If we introduce a new variable $u =R^3$, Eq. (\ref{near origin}) becomes
\begin{equation}
  2\pi\sigma u^{2/3}\exp\left(\frac{3i}{4\pi\sigma}\frac{\partial}{\partial u}\right)\Psi (u, \tau) =i\frac{\partial\Psi (u, \tau)}{\partial\tau}.
\end{equation}
The differential operator in the exponent is just a translation operator which shifts the argument of the wave function by a non-infinitesimal amount of $3i/(4\pi\sigma )$. Since the wave function is complex in general, a shift by a complex value is not a problem. Therefore Eq. (\ref{near origin}) can be written as
\begin{equation} \label{non-local}
  2\pi\sigma u^{2/3}\Psi\left(u+\frac{3i}{4\pi\sigma },\tau\right)=i\frac{\partial\Psi(u,\tau)}{\partial\tau}.
\end{equation}

The wave function near the origin $\Psi (R \rightarrow 0,\tau)$ is in fact related to the wave function at some distant point $\Psi (R \rightarrow (\frac{3i}{4\pi\sigma})^{1/3},\tau)$. This effect is non local and may have some implications for the information loss paradox \cite{Lowe:1999pk,Horowitz:2003he,Giddings:2006sj,Giddings:2006be}. At large distances far from the origin, non-local effects were absent (at least in the approximation we used) as can be seen from Eq. (\ref{appmom}) which gives a linear relation between the generalized velocity and the generalized momentum. However, in the last stages of the collapse, when $R \rightarrow 0$, these effects become important.  Eq. (\ref{no}) contains the generalized momentum (and thus the derivatives) in the exponent which makes the Schrodinger equation non-local. From Eq. (\ref{non-local}) we see that non-locality depends on the wall surface tension $\sigma$. For the light walls (small $\sigma$) non-locality is stronger, while large $\sigma$ suppresses it. However, the suppression can not be arbitrary large since the condition bounding $\sigma$ from above (given below Eq. (\ref{ISmass})) must be satisfied.

 While it is possible that the whole formalism breaks down at the very short distances of the order of Planck length, it is obvious that there will be a regime in which  non-local effects are important.

Eq. (\ref{non-local}) also implies that the wave function describing the collapsing shell is non-singular at the origin. Indeed, in the limit of $R \rightarrow 0$, this equation becomes
\be \label{non-singular}
 \frac{\partial\Psi(R \rightarrow 0,\tau)}{\partial\tau}    =0
\ee
 where we used the fact that the wave function at some finite $R$, i.e. $\Psi (R \rightarrow (\frac{3i}{4\pi\sigma})^{1/3},\tau)$, is finite. From Eq. (\ref{non-singular}) then follows that $\Psi(R \rightarrow 0)= {\rm const}$. Non-singularity of the wave function describing the collapsing object at the origin is very important knowing that the origin represents the classical singularity and it is the source of most of the problems and paradoxes in black hole physics. Some consequences will be discussed in conclusions.

\section{Conclusions}
\label{conclusions}

We studied gravitational collapse of the spherically symmetric shell of matter represented by a thin spherical domain wall in the context of quantum mechanics. We employed the inherently quantum functional Schrodinger formalism, which can be readily incorporated into the Wheeler-de Witt formalism. We examined most of the cases of interest.

In the literature, there exist some arguments that quantum effects near the horizon might change what an asymptotic observer would see in gravitational collapse. In particular, it was argued that the quantum fluctuations can make the collapse time finite for a static outside observer.  However, it appears that, at least in the framework of the functional Schrodinger formalism, quantum effects do not change the conclusions of classical general relativity, i.e. it takes infinite time according to the asymptotic observer's clock for the collapsing shell to cross its own Schwarzschild  radius. This is the conclusion valid only in the absence of quantum radiation. Hawking radiation of course makes the lifetime of the black hole finite and asymptotic observers will see infalling matter accumulated onto the stretched horizon \cite{Price:1986yy,Susskind:1993if}, thermalized, and eventually re-emitted as part of the Hawking radiation.

Results from the point of view of an infalling observer (an observer who is falling together with the collapsing shell) are of special interest. We first investigated what an infalling observer would see when the shell is crossing its own Schwarzschild radius.  To do that, we explored quantum effects in near horizon limit for an infalling observer and showed that, in the absence of quantum radiation, classical conclusions remain true, i.e. horizon is no obstacle for an infalling observer.

Finally, we explored quantum effects near the origin (i.e. classical singularity) from the point of view of an infalling observer. There are two important effects to be mentioned. First, the wave function describing the collapsing shell is non-singular at the origin. This is in agreement with the expectation that quantization will rid
gravity of singularities, just as in atomic physics it got rid
of the singularity of the Coulomb potential which has an identical classical $1/r$ behavior. If the singularity at the origin is really erased, than most of the assumed properties of the black holes need to be re-thought. In particular, once we include quantum radiation, a collapsing object (or a black hole) will lose all of its energy in finite time. However, in the absence of the true singularity at the center, a horizon formed during the collapse can not be a true global event horizon. In other words, in the absence of the singularity, a ``black hole" may trap the light and other particles for some finite amount of time, but not forever (nor is the information lost down the singularity). This has profound implications for  black hole physics.

Second, the quantum equation that governs physics near the classical singularity seems to be non-local. The Schrodinger equation describing the collapsing object contains an infinite number of derivatives. The dynamics of the wave function at certain point near the origin depends on the value of the wave function at some distant point. While these non-local effects were absent at large distances far from the origin, they become unsuppressed in the near origin regime. Non-local effects we are finding in our approach may signal two things. It may be that it is a simple consequence of the fact that the Wheeler-de Witt (or functional Schrodinger) formalism is only an approximation of some more fundamental local theory. After all, effective low energy actions are often non-local. The other possibility is that the quantum description of the black hole physics requires inherently non-local physics. Answer to this question requires further investigation.

\begin{acknowledgments} The authors are grateful to T.~Vachaspati,
  for very useful discussions. This work was supported by the HEPCOS group at the Department of Physics, SUNY at Buffalo.
 \end{acknowledgments}

\end{document}